\documentclass[twoside]{ilcws08}
\usepackage[latin1]{inputenc}
\usepackage[dvips]{graphicx,epsfig,color}
\usepackage{wrapfig,rotating}
\usepackage{amssymb,amsmath,array}
\usepackage{subfigure}

\begin{document}
\title{
WIMP Searches at the ILC using a model-independent Approach} 
\author{Christoph Bartels$^{1,2}$ and Jenny List$^1$
\thanks{The authors acknowledge the support by DFG grant Li 1560/1-1.}
\vspace{.3cm}\\
1- DESY - FLC \\
Notkestr. 85, 22607 Hamburg - Germany
\vspace{.1cm}\\
2- Universit\"at Hamburg - Inst. f. Experimentalphysik \\
Luruper Chaussee 149, 22761 Hamburg - Germany\\
}

\maketitle

\begin{abstract}
In this note the ILC's capabilities for detecting WIMPs and measuring their properties are studied.
The expected signal cross section is derived in a model-independent way from the observed relic density of Dark
Matter. Signal events are detected by means of initial state radiation (ISR).
The study is performed with a full simulation of the ILD detector. The results show that WIMPs are 
observable at the ILC if their coupling to electrons is 
not too small ($\mathcal{O}(0.1)$). Their masses can be measured with a precision \mbox{of 1 to 2 GeV}. 
The accessible phase space can be increased significantly using polarised beams, especially 
if the positrons are polarised as well.
\end{abstract}

\section{Introduction}

Due to its clean experimental environment the International Linear Collider (ILC) offers
the possibility to look for Weakly Interacting Massive Particles (WIMPs) in a model-independent way. This has been pointed out
in~\cite{Bartels:2007cv,birkedal}. In this study the expected sensitivity to such a WIMP signal, the achievable 
mass resolution and the influence of beam polarisation are investigated using a full detector 
simulation of the International Large Detector (ILD) of the ILC.\\

\section{Model-independent WIMP production cross section}
\label{sec:XSec}
It is assumed that the cosmic Dark Matter (DM) component is only due to one new type of particle, and
that the relic density is determined by pair annihilation of WIMPs into SM particles.
These annihilations may produce an $e^+e^-$ pair ($\chi\chi\rightarrow e^+e^-$) with an unknown 
branching fraction $\kappa_e$, where $\kappa_e \geq 0.3$ is strongly motivated
by recent PAMELA results~\cite{PAMELA,Adriani:2008zr}.
The total annihilation cross section $\sigma_{an}$ is then
determined by the observed \mbox{DM density $\Omega_{DM}$}.
The annihilation cross section required to match the observed 
relic density is in the order of a few pb~\cite{birkedal}. 
Using crossing relations one can derive the expected cross section for the reverse
process, i.~e. $e^+e^-\rightarrow \chi\chi$, which could be observable at the ILC. The resulting cross section contains
as free parameters:
\begin{itemize}
\item
the $e^+e^-$ branching fraction $\kappa_e$
\item
the mass of the WIMP $M_{\chi}$ and its spin $S_{\chi}$
\item
the angular momentum $J_{0}$ of the dominant partial wave 
($J_{0}$ corresponds to the spin of the exchange particle in the 
annihilation (production) process).
\end{itemize}
Since the produced WIMPs leave the detector without any further 
interaction an additional photon from ISR is required ($e^+e^-\rightarrow \chi\chi\gamma$).
The main background is the radiative neutrino pair production 
$e^+e^-\rightarrow \nu\nu\gamma$. This reaction is mediated dominantly by t-channel
\mbox{$W$-exchange} if the center of mass energy is significantly above the $Z^0$-pole.
Therefore \mbox{the $\nu\nu\gamma$ background} is strongly reducible depending on
the polarisation state of the initial electrons and positrons.\\
In this model-independent approach no constraints are imposed on the helicity structure of the WIMP couplings
to electrons, which therefore enter the production cross section as a new free
parameter. We consider three different scenarios for the helicity structure of the
WIMP couplings:
\begin{itemize}
\item
the same as the SM charged current interaction, i.~e. only $\kappa(e^-_Le^+_R)$ is nonzero
\item
parity and helicity are conserved: $\kappa(e^-_Le^+_R) = \kappa(e^-_Re^+_L)$
\item
opposite to the SM charged current interaction, only $\kappa(e^-_Re^+_L)$ is nonzero.
\end{itemize}
In the last two cases a significant enhancement of the signal to background ratio
is expected if electrons (and positrons) beams are appropriatly polarised. 

\section{Software and reconstruction tools}
For the SM $\nu\nu\gamma$ background the SLAC SM {\tt Whizard} event sample~\cite{SLACSM}
is used. This sample is generated with the {\tt Whizard} event generator including 
beamstrahlung and beam energy spread from {\tt Guinea Pig} for nominal ILC beam parameters.
Each $e^+e^-\rightarrow \nu\nu\gamma$ event contains up to two
{\em additional} ISR photons. The sample size corresponds to
10 fb$^{-1}$ of luminosity at $\sqrt{s} = 500$ GeV. The background sample
is reweighted with respect to the energy and polar angle of the detected photon 
according to the WIMP production and background cross sections. The benefit of this method is that
a full signal sample can be obtained for all investigated combinations of cross section
parameters with only one simulation and reconstruction cycle, reducing the
amount of processing time significantly. Since the predicted signal cross section 
assumes only one photon to be radiated, the center of mass energy of the hard subprocess
is used for the application of weights.\\
The event sample is then fed into the full {\tt Mokka 06-06-p03} ILD detector simulation 
(detector model LDCPrime\_02Sc) with a 3.5 Tesla magnetic field, and reconstructed with 
{\tt MarlinReco} using the Pandora Particle Flow algorithm (PFlow)~\cite{Thomson:2008zz}.

\begin{figure}
\setlength{\unitlength}{1.0cm}
\begin{picture}(14.0, 4.0) 
\put(0.0,0.0){\includegraphics[width=0.45\linewidth]{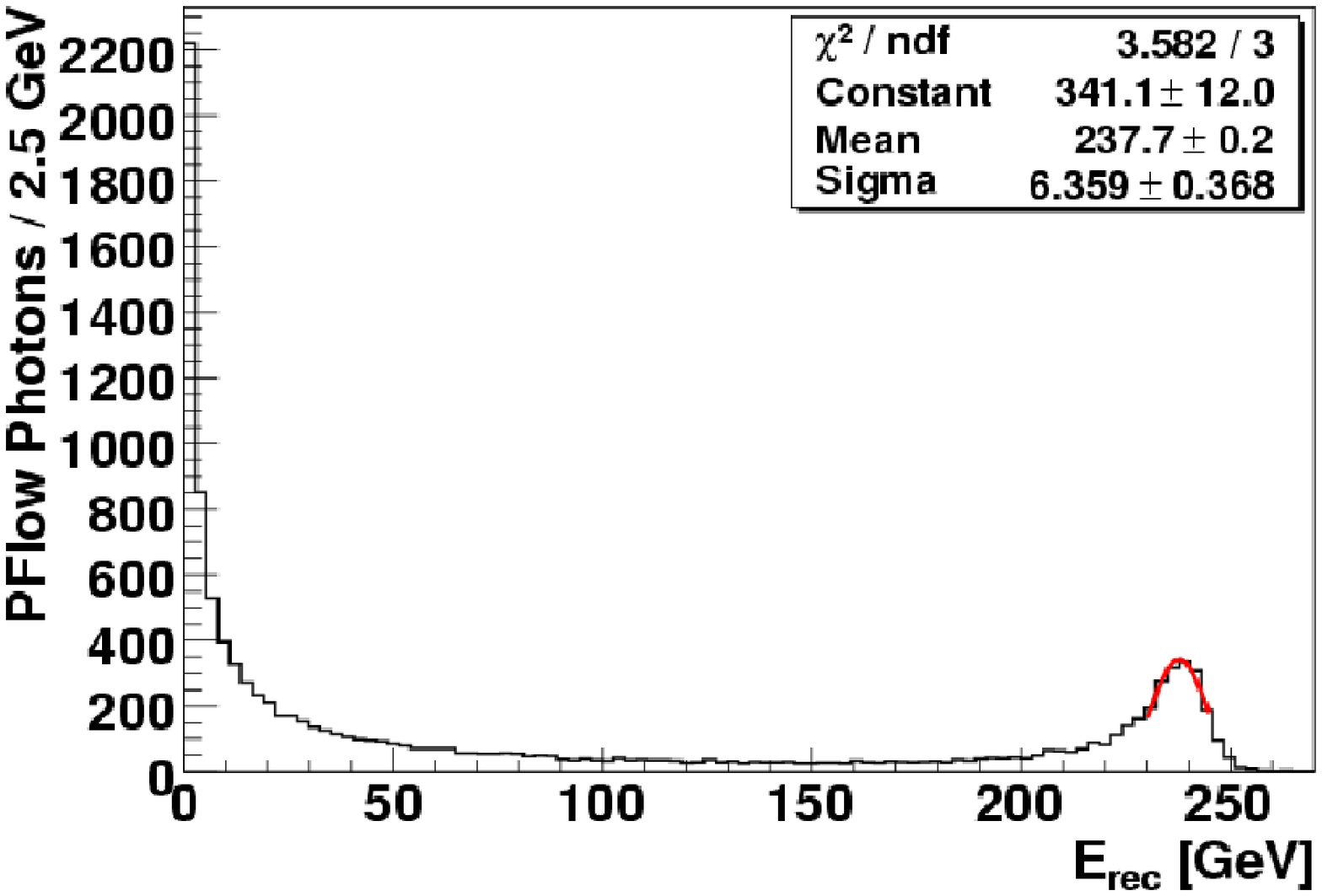}}
\put(7.0,0.0){\includegraphics[width=0.45\linewidth]{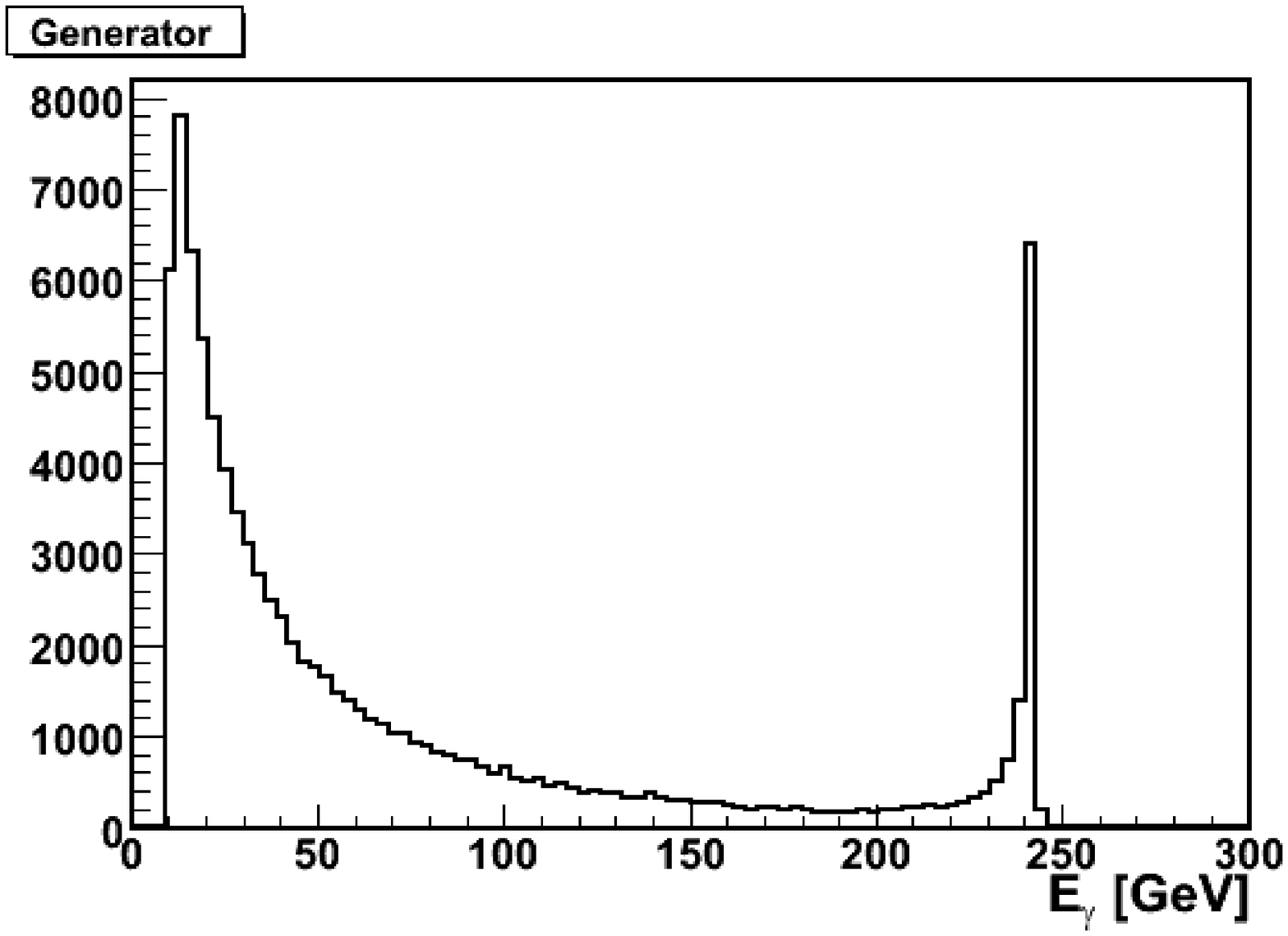}}
\put(0.0,0.0){(a)}
\put(7.0,0.0){(b)}
\end{picture}
\vspace{-0.5cm}
\caption{Energy spectrum of the most energetic photon of the $\nu\nu\gamma$ background 
(a) after reconstruction using Pandora PFlow and (b) on generator level.}
\label{Fig:gamma_rec}
\end{figure}

The event simulation and reconstruction process introduces a smearing of the reconstructed photon candidate
energies due to the intrinsic energy resolution of the detector and the performance
of the Particle Flow algorithm. Figure \ref{Fig:gamma_rec}(a) shows
the energy distribution of the most energetic photon candidate identified by the Pandora algorithm. 
The radiative return to the $Z^0$ at $\approx 240$ GeV is with a width of about 6.4 GeV significantly broadened
compared to the spectrum on generator level (Fig.~\ref{Fig:gamma_rec}(b)). 

\section{Cluster splitting}

\begin{figure}
\setlength{\unitlength}{1.0cm}
\begin{picture}(14.0, 4.0) 
\put(0.0,0.0){\includegraphics[width=0.45\linewidth]{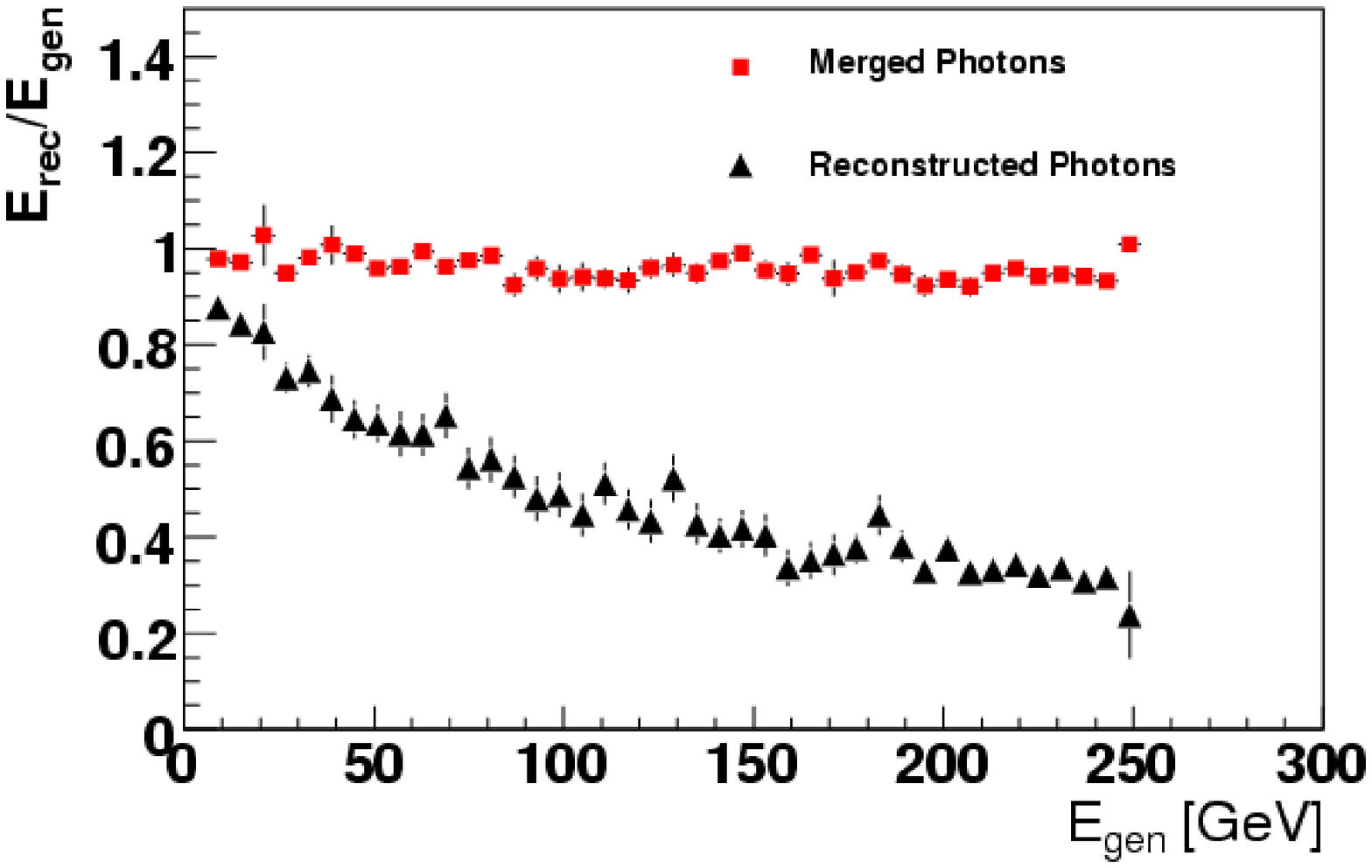}}
\put(7.0,0.0){\includegraphics[width=0.45\linewidth]{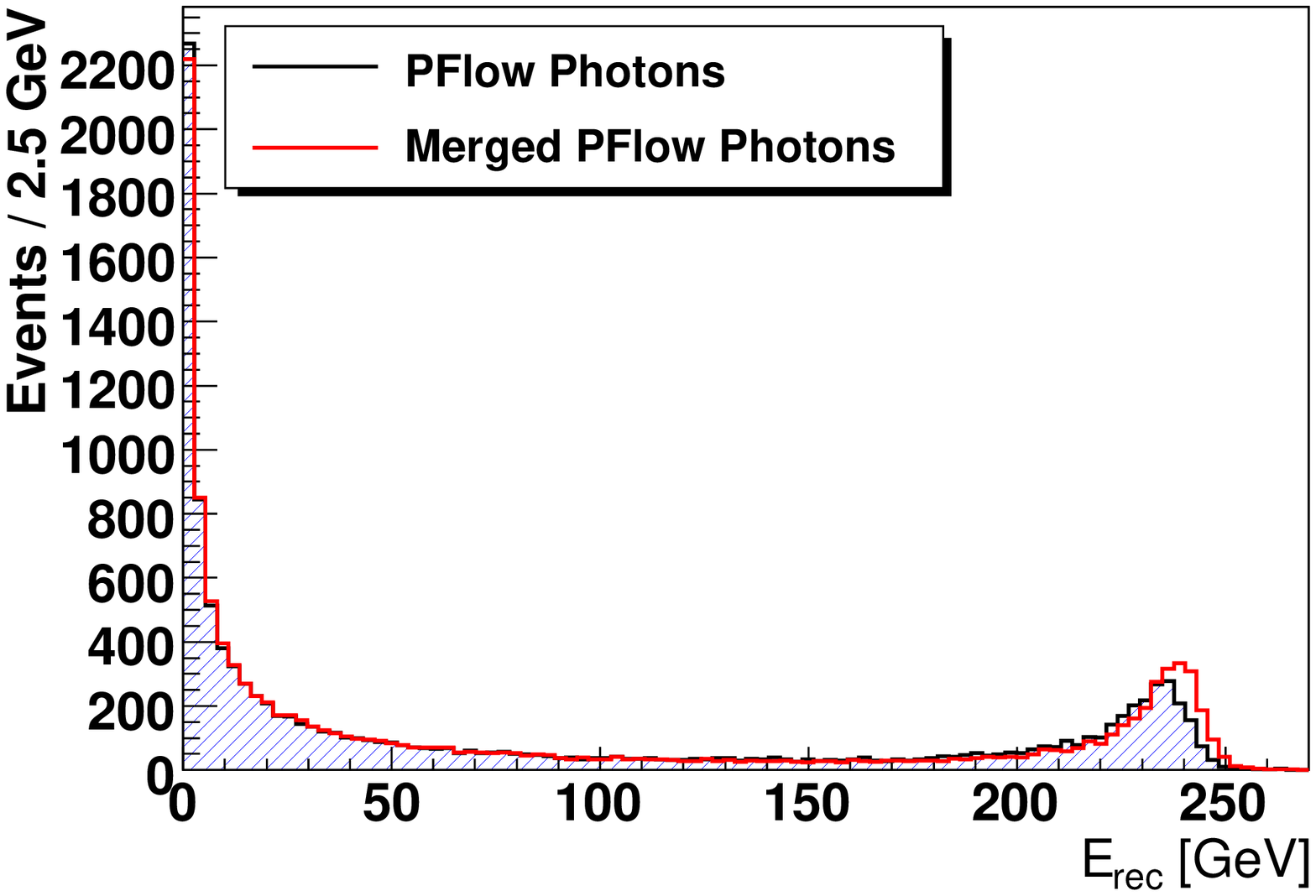}}
\put(0.0,0.0){(a)}
\put(7.0,0.0){(b)}
\end{picture}
\vspace{-0.5cm}
\caption{(a) Average ratio of reconstructed photon energy to generated photon energy before and after merging.
(b) Energy spectra before and after merging procedure.}
\label{Fig:gamma_ratio}
\end{figure}

The full reconstruction yields less photons at high energies
resulting in a shift of the $Z^0$ return from the expected central value of
241 GeV to 237.7 GeV (Fig.~\ref{Fig:gamma_rec}(a)). 
This behaviour is quantified in Figure \ref{Fig:gamma_ratio}(a) showing the average
ratio $E_{rec}/E_{gen}$ of the reconstructed photon candidate's energy to the photon energy on generator
level. The Pandora PFlow algorithm tends to split photon clusters stemming from one
generated photon into several photon candidates. This lowers the ratio $E_{rec}/E_{gen}$ especially
at high energies (black triangles). This effect can be compensated by applying a simple merging procedure.
Neighbouring photon candidates are combined to form a new set of photon candidates, with which
an average ratio of \mbox{$E_{rec}/E_{gen} \approx 1$} is recovered (red squares). 
The $Z^0$ return is also shifted back to its expected value (see Figure \ref{Fig:gamma_ratio}(b)). 

\section{Preliminary analysis results}
So far the following scenarios have been investigated:
\begin{itemize}
\item
WIMP spin: p-wave annihilation ($J_{0} = 1$) for $S_\chi = 1$ and $S_\chi = 1/2$
\item
WIMP couplings: $\kappa(e^-_Le^+_R) > 0, \kappa(e^-_Re^+_L) > 0$ and $\kappa(e^-_Le^+_R) = \kappa(e^-_Re^+_L) > 0$
\item
Polarisation: unpolarised beams, $e^-$ polarisation only ($P_{e^-} = 0.8$) and additional $e^+$ polarisation 
($P_{e^-} = 0.8$ and $P_{e^+} = 0.6$).
\end{itemize}

\subsection{Observational reach}
For each combination of these parameters the ILC reach for a 3$\sigma$ observation has been determined 
as a function of the WIMP mass $M_\chi$. The reach has been evaluated for an integrated luminosity of
500 fb$^{-1}$ at $\sqrt{s} = 500$ GeV using a larger event sample from a previous detector simulation~\cite{Bartels:2007cv}.
Due to the high SM neutrino production background,
the sensitivity has been obtained statistically by using fractional event counting~\cite{Junk:1999kv} as implemented 
in the {\tt ROOT} class TLimit.
\begin{figure}
\setlength{\unitlength}{1.0cm}
\begin{picture}(14.0, 2.5) 
\put(0.0,0.0){\includegraphics[width=0.33\linewidth]{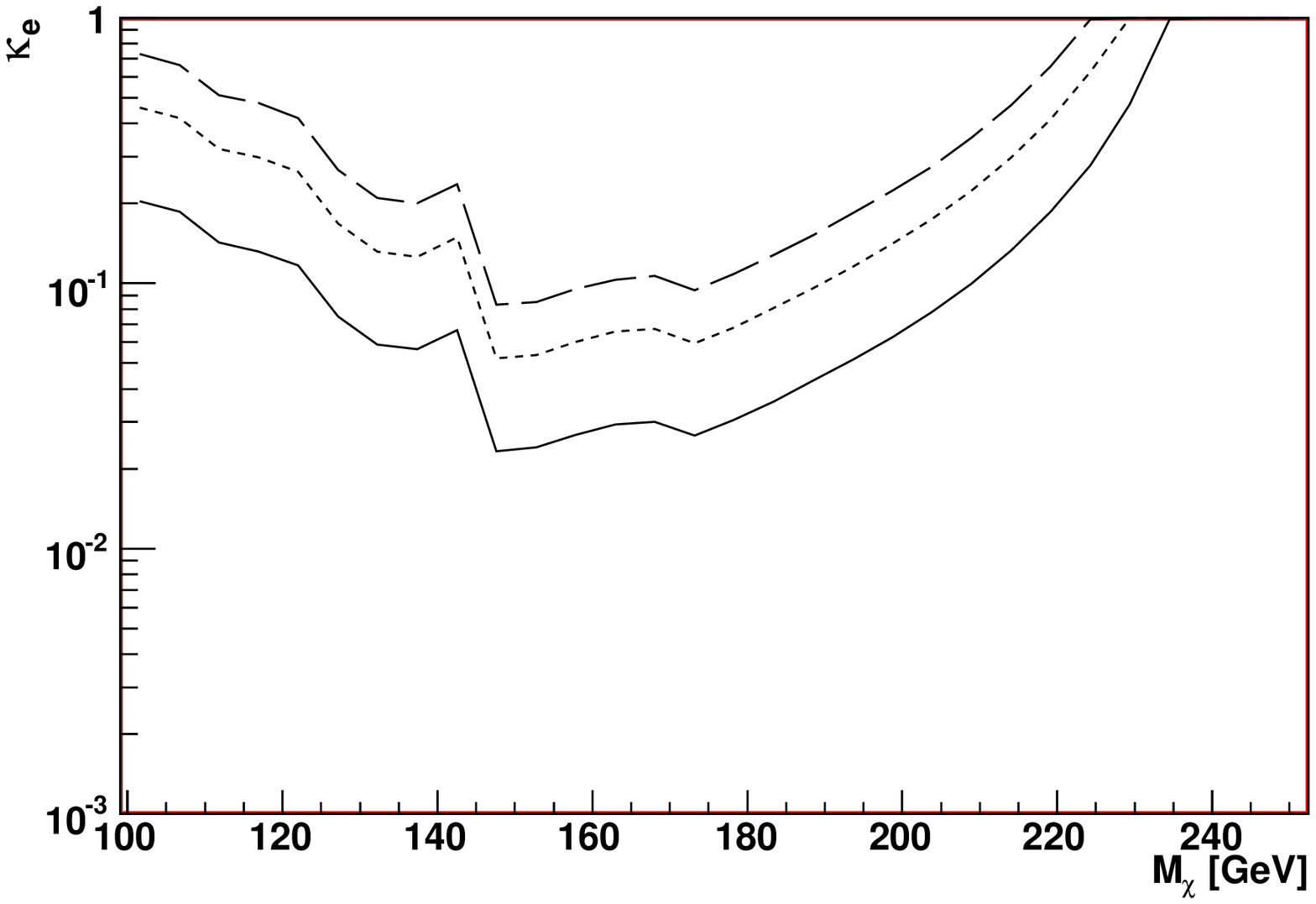}}
\put(5.0,0.0){\includegraphics[width=0.33\linewidth]{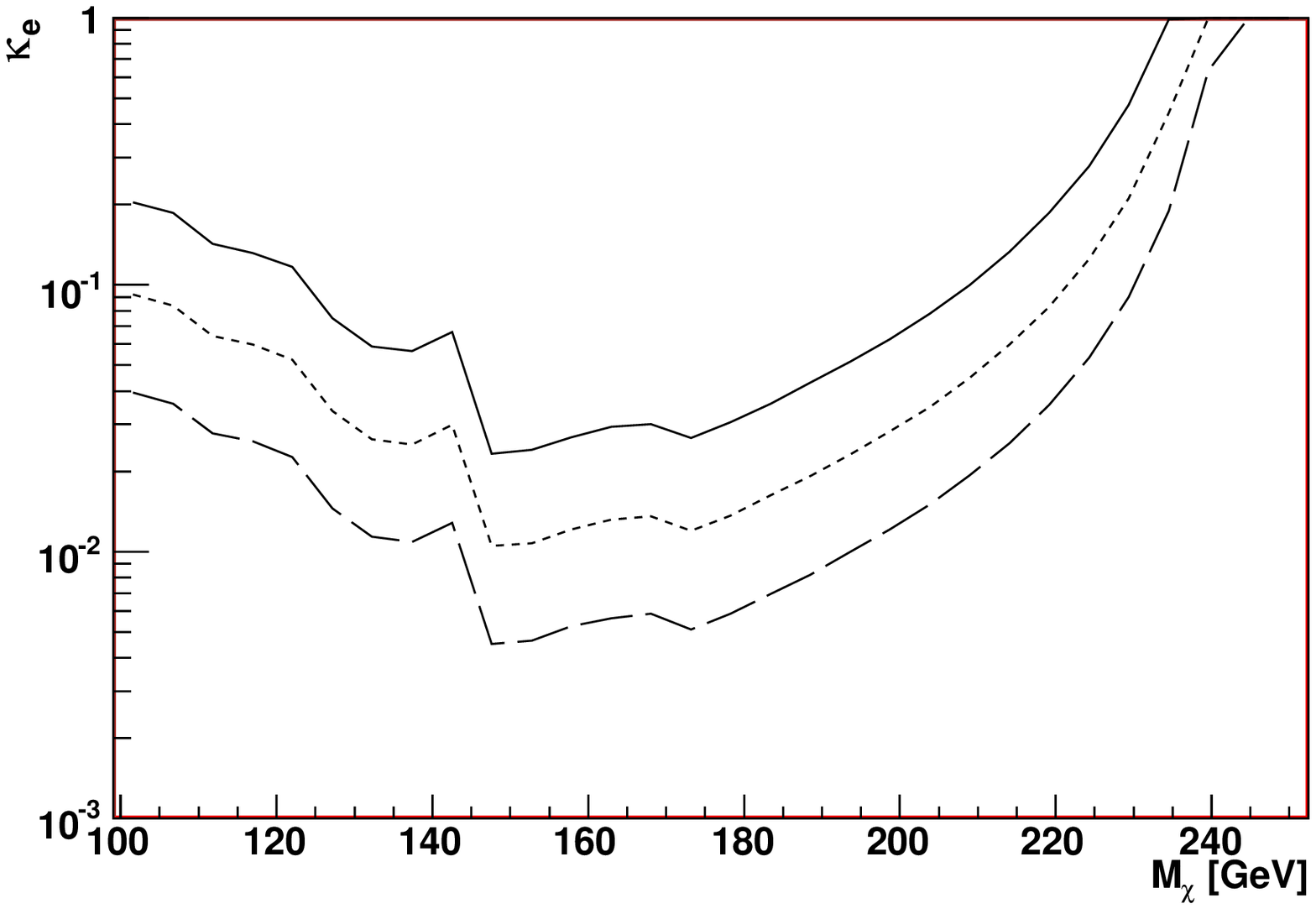}}
\put(10.0,0.0){\includegraphics[width=0.33\linewidth]{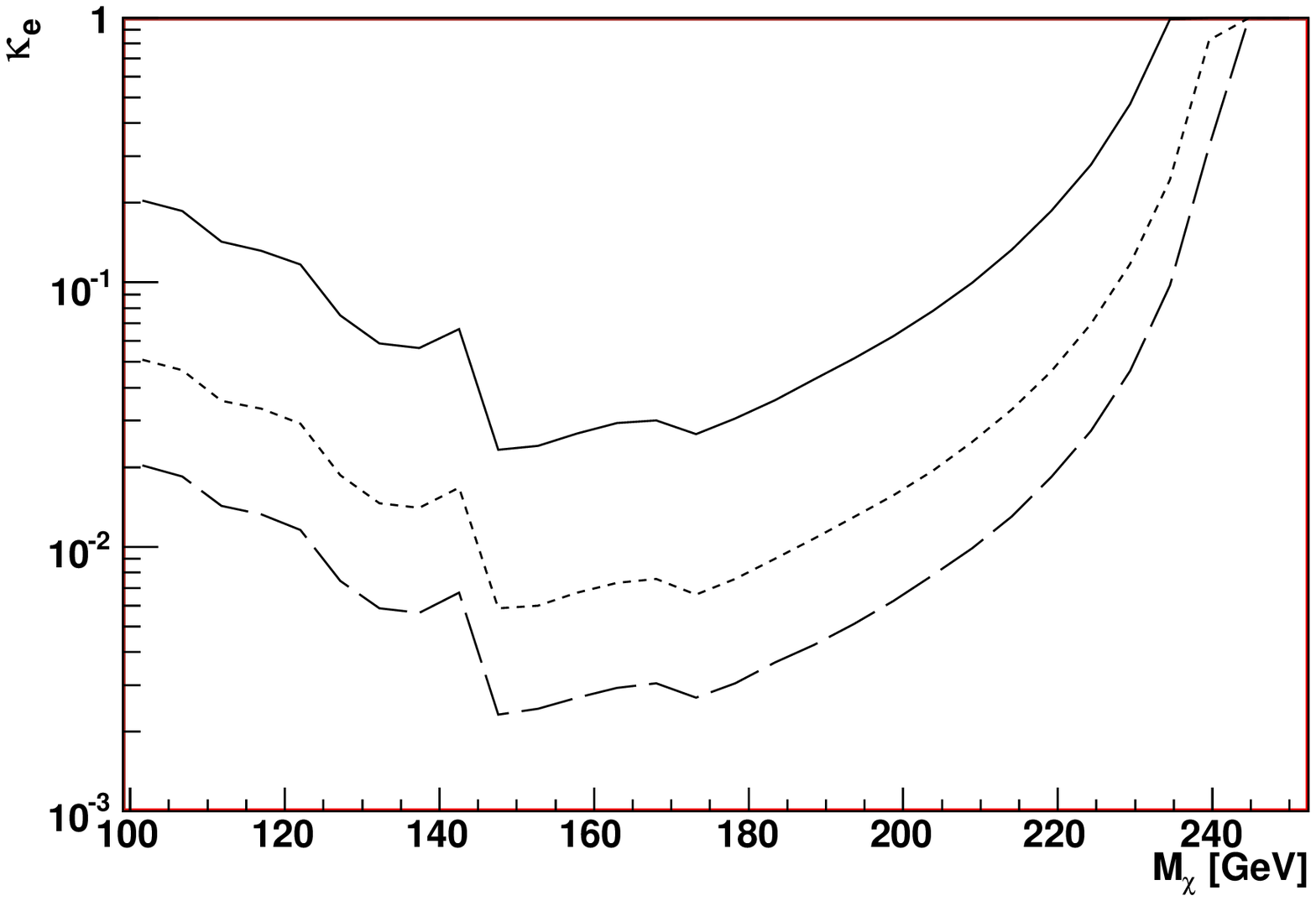}}
\put(0.0,-0.2){(a)}
\put(5.0,-0.2){(b)}
\put(10.0,-0.2){(c)}
\end{picture}
\vspace{-0.5cm}
\caption{Observational reach ($3\sigma$) of the ILC for a Spin-1 WIMP in terms of WIMP mass and $\kappa_e$ for three different chirality and couling scenarios. Full line: $P_{e^-} = P_{e^+} = 0$, dotted line: $P_{e^-} = 0.8, P_{e^+} = 0$, dashed line : $P_{e^-} = 0.8, P_{e^+} = 0.6$.}
\label{Fig:Case1}
\end{figure}
Figure~\ref{Fig:Case1} shows the expected ILC sensitivity for Spin-1 WIMPs in terms of the minimal observable branching fraction to electrons $\kappa_e$ as a function of the WIMP mass.  The regions above the curves are accessible, where
the full line gives the result for unpolarised beams, the dotted line for $P_{e^-} = 0.8$ and the dashed line for $P_{e^-} = 0.8$ and $P_{e^+} = 0.6$. Figure~\ref{Fig:Case1}(a) shows the case where the WIMPs couple only to lefthanded electrons and righthanded positrons, $\kappa(e^-_Le^+_R)$, Fig~\ref{Fig:Case1}(b) shows the parity and helicity conserving case, $\kappa(e^-_Le^+_R) = \kappa(e^-_Re^+_L)$, while Fig~\ref{Fig:Case1}(c) right plot is dedicated to the case that the WIMPs couple to righthanded electrons and lefthanded positrons ($\kappa(e^-_Re^+_L$)).

\begin{wrapfigure}{r}{0.5\columnwidth}
\vspace{-0.4cm}
\centerline{\includegraphics[width=0.45\columnwidth]{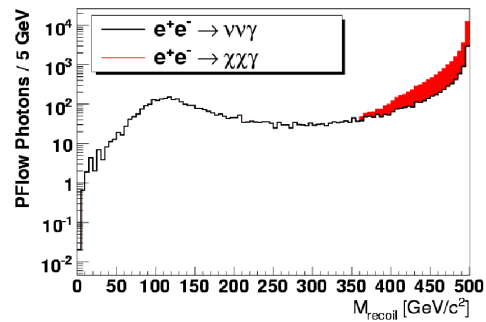}}
\vspace{-0.4cm}
\caption{Recoil mass distribution for a 180~GeV Spin-1 WIMP.}\label{Fig:Mass}
\end{wrapfigure}
For the latter two scenarios polarised beams increase the reach significantly, especially the additional positron polarisation increases the accessible range in $\kappa_e$ by about a factor of two. 
In case of a Spin-$\frac{1}{2}$ WIMP the sensitivity is somewhat worse, but again beam polarisation 
extends the observable part of the parameter space significantly~\cite{url}.
In all presented cases the ILC is sensitive to the branching \mbox{fraction $\kappa_e$ over} a large range of possible WIMP masses
and down to values below 0.3 indicated by the PAMELA results.

\subsection{Mass resolution}
\begin{figure}
\setlength{\unitlength}{1.0cm}
\begin{picture}(14.0, 3.5) 
\put(0.0,0.0){\includegraphics[width=0.45\linewidth]{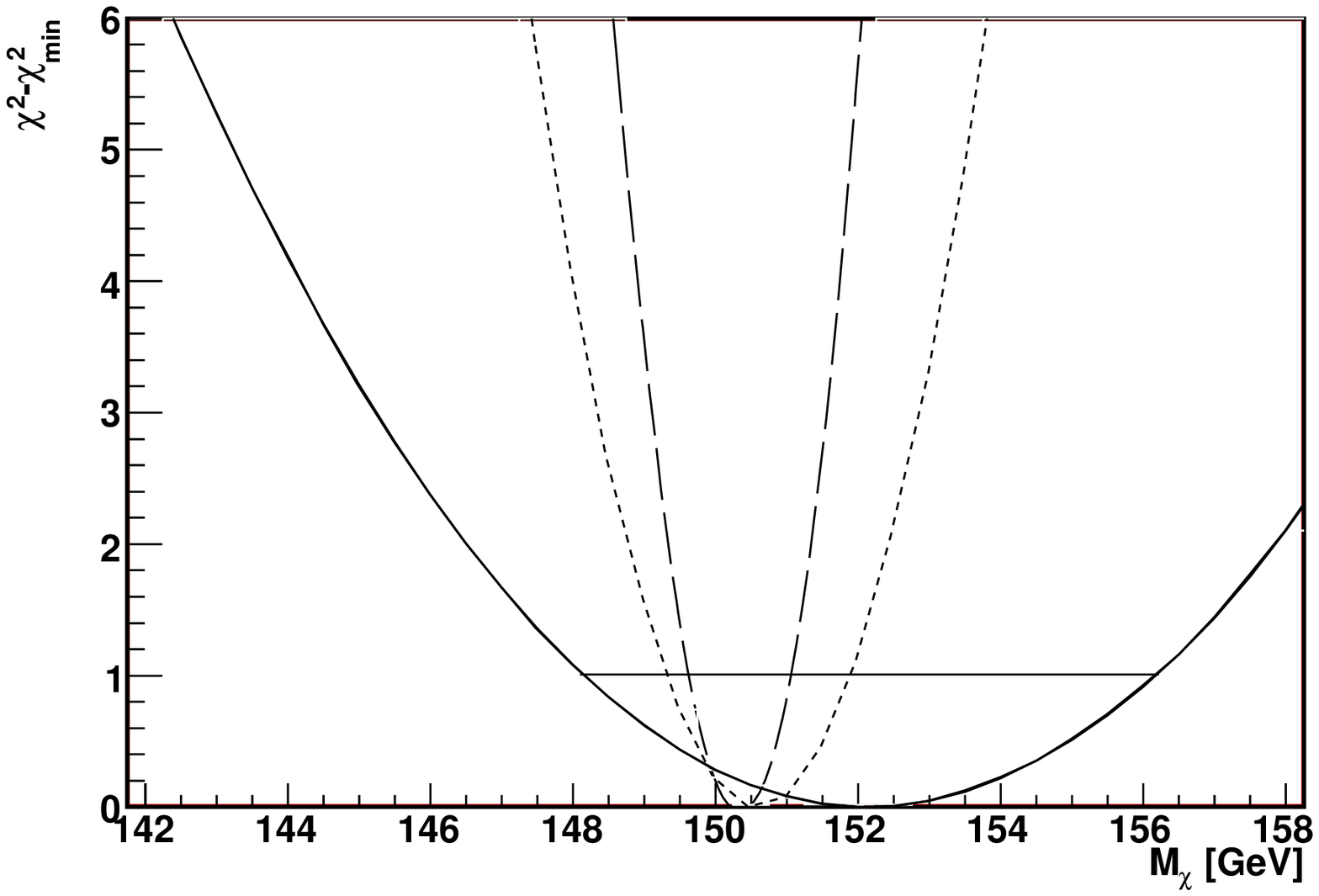}}
\put(7.0,0.0){\includegraphics[width=0.45\linewidth]{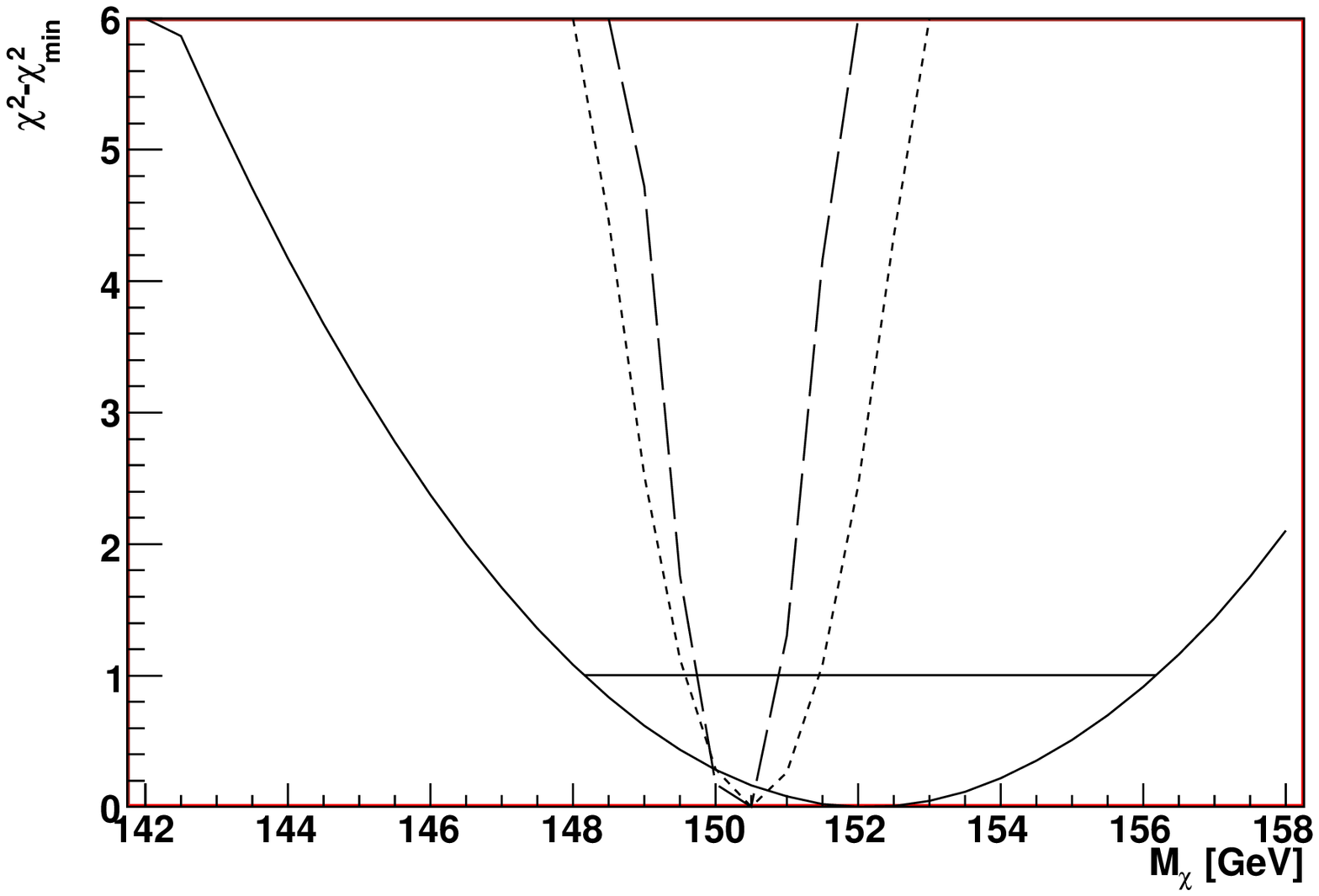}}
\put(0.0,0.0){(a)}
\put(7.0,0.0){(b)}
\end{picture}
\vspace{-0.5cm}
\caption{$\Delta \chi^2$ from mass determination by a template method for a Spin-1 WIMP with $M_X$=150~GeV. (a) parity and helicity conserving couplings, (b) $\kappa(e^-_Re^+_L)$. Full line: $P_{e^-} = P_{e^+} = 0$, dotted line: $P_{e^-} = 0.8, P_{e^+} = 0$, dashed line : $P_{e^-} = 0.8, P_{e^+} = 0.6$.}
\label{Fig:Mcase1}
\end{figure}

If WIMPs are observed at the ILC, their mass can be determined from the recoil mass distribution of the photons:
\begin{equation}
M_{\mathrm{recoil}}^2 = s -2\sqrt{s}E_{\gamma}
\end{equation}
Figure~\ref{Fig:Mass} shows an example for the recoil mass distribution for a 180~GeV \mbox{Spin-1} WIMP with both beams polarised. The WIMP signal shown in red (dark grey) on top of the SM neutrino background 
kicks in at $M_{\mathrm{recoil}} =$  360~GeV. From this distribution the WIMP mass can be 
reconstructed e.~g.~with a template method. 
For this procedure, only 200~fb$^{-1}$ of the available MC sample have been analysed as dataset, 
the rest is used for the templates. Figure~\ref{Fig:Mcase1} shows the obtained $\Delta\chi^2$ as 
function of the reconstructed WIMP mass for a 150~GeV \mbox{Spin-1} WIMP for $\kappa_e = 0.3$. 
Fig.~\ref{Fig:Mcase1}(a) shows the helicity and parity conserving case, while (b) illustrates the case 
that the WIMPs couple to righthanded electrons and lefthanded positrons, $\kappa(e^-_Re^+_L)$. 
Again the full line gives the result for unpolarised beams, the dotted line for $P_{e^-} = 0.8$ and 
the dashed line for $P_{e^-} = 0.8$ and $P_{e^+} = 0.6$. Without any beam polarisation, the mass 
resolution is about 4~GeV, which is reduced to about 1.2~GeV by switching on the electron polarisation. 
Additional positron polarisation improves the resolution by another factor of two to about 0.6~GeV.

Similar results are obtained for Spin-$\frac{1}{2}$ WIMPs~\cite{url}.
As for the observation reach, the situation is slightly worse than in the Spin-1 case, but again the use of beam polarisation leads to a significant gain in resolution.

\section{Conclusions and outlook}

This study is one example of a physics analysis using the full simulation
of the ILD detector concept. It is part of the optimisation effort for the ILD detector,
and is further improved in parallel to the ongoing optimisation of the detector model.
It demonstrates that there is a good chance of detecting WIMPs at the ILC in a model-independent
way, and to measure their mass with a precision of about 1 GeV.
The preliminary results show that the reach of the ILC on the branching fraction $\kappa_e$ of
WIMP pair annihilation to electrons covers values of $\kappa_e$ down to $\simeq 0.01$, which is well below the $\kappa_e \geq 0.3$ 
indicated by the recent PAMELA data. Both the range in phase space, as well as the mass resolution
improve significantly when polarsised beams are assumed. Typically the use of 80\% electron polarisation
gives improvements of a factor of two over unpolarised beams, whereas an additional positron 
polarisation of 60\% yields another factor of two.\\
The results will be updated using different Particle Flow algorithms 
and Photon finders. Reducible backgrounds as well as beamstrahlung will be included. 
The obtained results will also be compared to specific SUSY
scenarios in which the only open SUSY channel at the ILC's center of mass energy is
radiative neutralino production~\cite{Dreiner:2007vm}.


\begin{footnotesize}



%

\end{footnotesize}


\end{document}